\documentclass[prd,aps,a4paper,twocolumn,nofootinbib]{revtex4-1}  

\newif\ifusesec
\usesectrue  
   
\usepackage{graphicx} 
\usepackage{mathrsfs}
\usepackage{amsmath,amsfonts,amssymb}
\usepackage{multirow}

\newcommand{\beq}{\begin{equation}}
\newcommand{\eeq}{\end{equation}}

\begin{document}

\title{Confirming and improving post-Newtonian and effective-one-body results from self-force computations along eccentric orbits around a Schwarzschild black hole}

\author{Donato \surname{Bini}$^1$}
\author{Thibault \surname{Damour}$^2$}
\author{Andrea \surname{Geralico}$^1$}

\affiliation{$^1$Istituto per le Applicazioni del Calcolo ``M. Picone'', CNR, I-00185 Rome, Italy\\
$^2$Institut des Hautes Etudes Scientifiques, 91440 Bures-sur-Yvette, France}

\date{\today}

\begin{abstract}
We analytically compute, through the six-and-a-half post-Newtonian order, the second-order-in-eccentricity piece of the Detweiler-Barack-Sago gauge-invariant redshift function for a small mass in eccentric orbit around a Schwarzschild black hole. Using the first law of mechanics for eccentric orbits [A. Le Tiec, Phys. Rev. D  {\bf 92}, 084021 (2015)] we transcribe our result into a correspondingly accurate knowledge of the second radial potential of the effective-one-body formalism [A. Buonanno and T. Damour, Phys. Rev. D {\bf 59}, 084006 (1999)]. We compare our newly acquired analytical information to several different numerical self-force data and find good agreement, within estimated error bars. We also obtain, for the first time, independent analytical checks of the recently derived, comparable-mass fourth-post-Newtonian order dynamics [T. Damour, P. Jaranowski and G. Shaefer, Phys. Rev. D {\bf 89}, 064058 (2014)].
\end{abstract}

\pacs{04.20.Cv, 98.58.Fd}
\keywords{eccentric orbits, black holes}
\maketitle

\section{Introduction}
Recent years have witnessed a useful {\it synergy} between various ways of tackling, in General Relativity, the two-body problem. In particular, the effective-one-body (EOB) formalism \cite{Buonanno:1998gg, Buonanno:2000ef, Damour:2000we, Damour:2001tu} has served as a focal point allowing one to gather and compare  information contained in various other approaches to the two-body problem, such as post-Newtonian (PN) theory \cite{Schafer:2009dq,Blanchet:2013haa}, self-force (SF) theory \cite{Barack:2009ux,Poisson:2011nh}, as well as full numerical relativity simulations.

The aim of the present work is to extract new information on the dynamics of {\it eccentric} (non-spinning) binary systems from both analytical and numerical SF computations along eccentric orbits around a Schwarzcshild black hole. This new information will concern both the usual PN-expanded approach to binary systems, and its EOB formulation (which, as we shall see, is particularly useful for transforming the information between various gauge-invariant observable quantities).

The first gauge-invariant quantity we shall consider is the generalization to eccentric orbits of Detweiler's \cite{Detweiler:2008ft} inverse redshift function, namely the function
\beq
\label{I1}
U\left(m_2\Omega_r, m_2\Omega_\phi, \frac{m_1}{m_2}\right)= \frac{\displaystyle\oint dt}{\displaystyle\oint d\tau}=\frac{T_r}{{\mathcal T}_r}
\eeq
introduced by Barack and Sago \cite{Barack:2011ed}.

The notation here is as follows. The two masses of the considered binary system are $m_1$ and $m_2$, with the convention $m_1\le m_2$ (and $m_1\ll m_2$ in SF calculations). We then denote (in our EOB considerations) $M\equiv m_1+m_2$, $\mu\equiv m_1m_2/(m_1+m_2)$, $\nu\equiv \mu/M=m_1m_2/(m_1+m_2)^2$. The intensity of the gravitational potential is measured (in EOB theory) by $u=M/r$ (in  the units $G=c=1$ we use). In Eq. \eqref{I1} the symbol $\oint $ denotes an integral over a radial period (from periastron to periastron) so that $T_r=\oint dt$ denotes the coordinate-time period and ${\mathcal T}_r=\oint d\tau$ the proper-time period. In addition $\Omega_r=2\pi/T_r$ is the radial frequency and $\Omega_\phi=\oint d\phi/\oint dt=\Phi/T_r$ is the mean azimuthal frequency. The first-order SF contribution $\delta U$ to the function \eqref{I1}, defined by 
\begin{eqnarray}
U\left(m_2\Omega_r, m_2\Omega_\phi, \frac{m_1}{m_2}\right)&=&U_0\left(m_2\Omega_r, m_2\Omega_\phi\right)\nonumber\\
&+& \frac{m_1}{m_2}\delta U\left(m_2\Omega_r, m_2\Omega_\phi\right)\nonumber\\
&+& O\left( \frac{m_1^2}{m_2^2} \right)\,,
\end{eqnarray}
is conveniently represented as a function of the dimensionless semi-latus rectum $p$ and the eccentricity $e$ of the unperturbed orbit: $\delta U(p,e)$. [See below for explicit definitions.]

Our first result will be to analytically derive the eccentricity-dependent part (denoting $u_p\equiv 1/p$) of
\begin{eqnarray}
\delta U(p,e)&=& \delta U^{e^0}(u_p)+e^2 \delta U^{e^2}(u_p)\nonumber\\
&+& e^4 \delta U^{e^4}(u_p)+e^6 \delta U^{e^6}(u_p)+O(e^8)
\end{eqnarray}
to order $O(u_p^{15/2})$ included for the $O(e^2)$ piece $\delta U^{e^2}$, and to order $u_p^5$ included for the $O(e^4)$ piece.
The previous knowledge of the coefficients $\delta U^{e^2}$, $\delta U^{e^4}$, $\delta U^{e^6}$ was only $O(u_p^4)$, corresponding to the 3PN level \cite{Tiec:2015cxa}. Then we shall translate our higher-order results on  $\delta U^{e^2}$ into a correspondingly improved result (6.5PN level)  for the EOB potential $\bar d(u)$ entering the dynamics of eccentric orbits at the $p_r^2$ level.
To do this we shall use a recent generalization to eccentric orbits, \cite{Tiec:2015cxa}, of the connection between $\delta U^{e^0}(u_p)$ and the $O(\nu)$ piece $a(u)$ of the main EOB radial potential \cite{Barausse:2011dq,Akcay:2012ea}. Let us note in passing that this connection has a direct link with what was, historically, the starting point of the EOB formalism, i.e., the (gauge-invariant) \lq\lq action-angle" (Delaunay) form of the two-body Hamiltonian \cite{Damour:1988mr}.
Using the relations derived in \cite{Tiec:2015cxa} will allow us to provide, among other results, the first explicit checks (done by a completely different analytical approach) of the recently derived (comparable-mass) 4PN dynamics \cite{Jaranowski:2013lca,Bini:2013zaa,Damour:2014jta,Jaranowski:2015lha,Damour:2015isa}.

In addition, we will use  our improved results for performing several different comparisons  with (and information-extraction from) various SF numerical data on eccentric orbits \cite{Barack:2010ny,Barack:2011ed,Akcay:2012ea,Akcay:2015pza,vandeMeent:2015lxa}.

Let us finally anticipate our conclusions by recalling that the first work suggesting  several explicit ways of extracting information of direct meaning for the conservative dynamics of comparable-mass systems (especially when formulated within the EOB theory) \cite{Damour:2009sm} has pointed out other gauge-invariant observables which have not yet been explored by the SF community but which offer, as significant advantage over the presently explored \lq\lq eccentric redshift" observable, the possibility of probing more deeply into the strong-field regime. Indeed, as we shall discuss below, the expansion of $\delta U(p,e)= \delta U^{e^0}(u_p)+e^2 \delta U^{e^2}(u_p)+\ldots$ encounters a singularity at the last stable (circular) orbit (LSO) $u_p=1/p=1/6$ which prevents \footnote{One should, however, note that if one does not expand $\delta U(p,e)$ in powers of $e$, one can, in principle, be sensitive to the EOB potentials up to $u_p=\frac14$ corresponding to the marginally bound motion with $e=1$ and $p=4$.} one for using current SF calculations on eccentric orbits to explore the domain $u_p\ge \frac16$.
By contrast, some of the gauge-invariant observables described in \cite{Damour:2009sm}  allow one, in principle, to explore the $O(\nu)$ EOB potentials up to  $u=\frac13$ (corresponding to the Schwarzschild light-ring).

\section{High PN-order analytical computation of the self-force correction to the averaged redshift function along eccentric orbits}

Barack and Sago \cite{Barack:2011ed} have introduced a generalization to eccentric orbits of Detweiler's \cite{Detweiler:2008ft} gauge-invariant first-order SF correction
to the (inverse) redshift. 
This gauge-invariant measure of the $O(m_1/m_2)$ conservative SF effect on eccentric orbits is denoted as $\delta U(m_2\Omega_r, m_2\Omega_\phi)$. It is  a function of the two $m_2$-adimensionalized  fundamental frequencies of the orbit, $\Omega_r=2\pi/T_r$ and $\Omega_\phi=\Phi/T_r$ where $T_r$ is the radial period and $\Phi$ the angular advance during one radial period.
It is given in terms of the $O(m_1/m_2)$ metric perturbation $h_{\mu\nu}$, where 
\beq
g_{\mu\nu}(x^\alpha; m_1, m_2)=g^{(0)}_{\mu\nu}(x^\alpha; m_2)+\frac{m_1}{m_2} h_{\mu\nu}(x^\alpha)+O\left( \frac{m_1^2}{m_2^2} \right)
\eeq
[with $g^{(0)}_{\mu\nu}(x^\alpha; m_2)$ being the Schwarzschild metric of mass $m_2$] by the following
 time average
\beq
\label{delta_U1}
\delta U (p,e)=\frac12 \, (U_{0})^2\langle h_{uk}\rangle_{t}\,.
\eeq
Here, we have expressed $\delta U$ (which is originally defined as a {\it proper} time $\tau$ average \cite{Barack:2011ed}) in terms of the {\it coordinate} time $t$ average of the mixed contraction $h_{uk}=h_{\mu\nu}u^\mu k^\nu$ where $u^\mu\equiv u^t k^\mu$, $u^t=dt/d\tau$ and $k^\mu\equiv \partial_t +dr/dt\partial_r +d\phi/dt \partial_\phi$. [Note that in the present eccentric case the so-defined $k^\mu=u^\mu/u^t$ is no longer a Killing vector.] In Eq. \eqref{delta_U1} we considered $\delta U$ as a function of the dimensionless semi-latus rectum $p$ and eccentricity  $e$ (in lieu of $m_2\Omega_r$,  $m_2\Omega_\phi$) of the {\it unperturbed} orbit, as is allowed in a first-order SF quantity. In addition, $U_0$ denotes the proper-time average of $u^t=dt/d\tau$ along the unperturbed orbit, i.e., the ratio $U_0={T_r}/{{\mathcal T}_r}|_{\rm unperturbed}$.
The quantities $p$ and $e$ are defined  by writing   the minimum (pericenter,
$r_{\rm peri}$) and maximum (apocenter, $r_{\rm apo}$) values of
the Schwarzschild  radial coordinate along an (unperturbed) eccentric orbit as 
\beq
r_{\rm peri}=\frac{m_2 p}{1+e}\,,\qquad r_{\rm apo}=\frac{m_2 p}{1-e}\,.
\eeq
They are in correspondence with the conserved (dimensionless) energy $E=-u_t$ and angular momentum $L= u_\phi/m_2$ of the background orbit, via
\beq
\label{E_and_L_in_p_e}
E^2= \frac{ (p-2)^2-4e^2  }{p(p-3-e^2)}\,,\qquad L^2=\frac{p^2 }{p-3-e^2}\,.
\eeq
The domain of the $p$-$e$ plane parametrizing bound eccentric orbits is defined by
\beq
\label{separatrix}
p> 6 +2e\,,\qquad e< 1\,.
\eeq
As is well known, the values  of the  frequencies $\Omega_{r 0}$ and $\Omega_{\phi 0}$ along an unperturbed eccentric orbit, as well as the periastron advance $\Phi_0=2\pi K=2\pi (1+k)$ (in the notation of \cite{damour_deruelle1,damour_deruelle2}), the proper-time radial period ${\mathcal T}_{r0}=\oint d\tau$  and therefore $U_0=T_{r0}/{\mathcal T}_{r0}$, are expressible in terms of elliptic integrals. For instance, 
\begin{eqnarray}
\label{Phi0}
\Phi_0&=&2\pi K=\oint d\phi\nonumber\\
&=&4\sqrt{\frac{p}{p-6-2e }}{\rm EllipticK}\left[k^2=\frac{4e}{p-6-2e}\right]\,,
\end{eqnarray}
where  EllipticK is a complete elliptic integral. Though it is not manifest in Eq. \eqref{Phi0}, $\Phi_0$ (as well as the other above-mentioned quantities)   is an even function of $e$, as e.g., exhibited in Eq. A.8 of \cite{Damour:1988mr}.

The correction $\delta U $ is equivalent to the correction $\delta z_1$ to the (coordinate-time) averaged redshift $z_1$
\beq
z_1=\left\langle \frac{d \tau}{dt}\right\rangle_t = \left(\left\langle \frac{dt}{d \tau}\right\rangle_\tau \right)^{-1}=U^{-1}\,,
\eeq
namely
\beq
\label{eq_z1}
\delta z_1=-\frac{\delta U}{U_0^2}=-\frac12 \langle h_{uk}\rangle_{t}\,.
\eeq
 
We have analytically computed $\delta U(p,e)$ at second order in eccentricity   and up to order $O(1/p^{15/2})$, which corresponds to the 6.5PN order.
Our computation is based on an extension of the technology we used in our previous papers, see notably \cite{Bini:2013zaa,Bini:2013rfa}. The crucial modification that we needed to tackle in the present {\it eccentric} analytical calculation was the existence of two orbital frequencies $\Omega_{r 0}$ and $\Omega_{\phi 0}$ in the motion. As a consequence, the nine (original\footnote{Before their transformation into odd and even  source-terms of a Regge-Wheeler equation.}) source terms  in the Regge-Wheeler-Zerilli equations have a structure of the type $f(r)e^{im \phi_0(t)}\delta (r-r_0(t))$ that must be   
evaluated along the unperturbed particle motion  $r_0(t)$, $\phi_0(t)$. 

Up to order $e^2$ included,  the motion is explicitly given by 
\begin{eqnarray}
\frac{r_0(t)}{ m_2\, p}&=&
\frac{1}{1-e} +e(\cos \Omega_{r 0} t-1)\nonumber\\
&+&e^2 b_2(p)(\cos(2\Omega_{r 0} t)-1)+O(e^3)\nonumber\\
\phi_0(t)  
&=&  \Omega_{\phi 0} t+ ec_1(p)\sin(\Omega_{r 0} t)\nonumber\\
&+&e^2 c_2(p)\sin(2\Omega_{r 0} t) +O(e^3)\,,
\end{eqnarray}
where
\begin{eqnarray}
c_1(p)&=& -\frac{2 (p-3) }{ (p-2) }\left(1-\frac{6}{p}\right)^{-1/2}\nonumber\\
c_2(p)&=& \frac{(5 p^3-64 p^2+250 p-300) }{4(p-6) (p-2)^2} \left(1-\frac{6}{p}\right)^{-1/2}\nonumber\\
b_2(p)&=& -  \frac{(p^2-11 p+26)}{ 2(p-2) (p-6)}\,.\\
& & \phantom{X}\nonumber
\end{eqnarray}

Note that one could  conveniently   express both $r$ and $\phi/K=2\pi  {\phi}/{\Phi_0}$
as periodic functions of the \lq\lq mean anomaly"   $\ell =\Omega_{r 0} t$. [The time origin is chosen so that $t=0$ (and $\ell=0$, modulo $2\pi$) corresponds to an apoastron.]

The expansion of the source-terms (which originally contain $\delta(r-r_0(t))$ and at most two of its derivatives) in powers of $e$ generates, at order $e^2$, up to four derivatives of $\delta(r-m_2/p)$ in the even part and up to three in the odd part.
This expansion gives rise to multiperiodic coefficients in the source terms, involving the combined frequencies
\beq
\omega_{m,n}=m\Omega_{\phi 0}+n \Omega_{r 0}
\eeq
with $n=0,\pm 1, \pm 2$ when working as we do up to order $e^2$.

For the present computation we have used, for the Green function, the Mano-Suzuki-Takasugi \cite{Mano:1996mf,Mano:1996vt} hypergeometric expansions up to multipolar order $l=4$ and our PN-expanded solution for $l>4$. 
A feature of our formalism is that, in order to compute the {\it regularized} value of $\langle h_{uk}\rangle_t$, we do not need
to analytically determine in advance the corresponding subtraction term, because we automatically obtain it as a side-product of our computation [by taking the 
$l\to \infty$ limit of our PN-based calculation]. The expansion in powers of $u_p\equiv 1/p$ of the constant $B$ to be subtracted from $\delta U$  is found to be
\begin{widetext}
\begin{eqnarray}
B&=&  2 u_p-\frac12 u_p^2-\frac{39}{32}u_p^3-\frac{385}{128}u_p^4-\frac{61559}{8192}u_p^5-\frac{622545}{32768}u_p^6-\frac{25472511}{524288}u_p^7-\frac{263402721}{2097152}u_p^8 \nonumber\\
&+&  \left(-2 u_p+\frac74 u_p^2+7 u_p^3+\frac{8597}{256}u_p^4+\frac{1498513}{8192}u_p^5+\frac{69481763}{65536}u_p^6+\frac{1650414477}{262144} u_p^7+\frac{158088550401}{4194304}u_p^8 \right)e^2\nonumber\\
&& +O(u_p^9,e^3)\,.
\end{eqnarray}
\end{widetext}

As usual the low multipoles ($l=0,1$) have been computed separately, as in Eq. (138) of Ref. \cite{vandeMeent:2015lxa}. The corresponding (already subtracted) contribution to $\delta U$ is the following
\begin{widetext}
\small{
\begin{eqnarray}
\delta U^{l=0,1}&=&-2 u_p+2 u_p^2+\frac{3}{16} u_p^3-\frac{695}{64} u_p^4-\frac{240841}{4096} u_p^5-\frac{3949743}{16384} u_p^6-\frac{233188353}{262144} u_p^7 -\frac{3259311903}{1048576} u_p^8\nonumber\\
&+&\left(2 u_p-\frac32 u_p^2-\frac{29}{4} u_p^3-\frac{7317}{128} u_p^4-\frac{1483601}{4096} u_p^5-\frac{67773219}{32768} u_p^6-\frac{1501264013}{131072} u_p^7-\frac{133483493377}{2097152} u_p^8\right) e^2\nonumber\\
&&+O(u_p^9,e^3)\,.
\end{eqnarray}
}
\end{widetext}

Our final result reads
\begin{eqnarray}
\label{deltaU_PN}
\delta U(p,e)&=&\delta U^{e^0}(u_p)+e^2 \delta U^{e^2}(u_p)+ e^4 \delta U^{e^4}(u_p)\nonumber\\
&+& e^6 \delta U^{e^6}(u_p)+O(e^6)\,.
\end{eqnarray}
Here  $\delta U^{e^0}(u_p)$ is the circular orbit Schwarzschild SF result which has been determined to very high PN accuracy in previous works \cite{Bini:2015bla,Kavanagh:2015lva},  
$\delta U^{e^2}(u_p)$ is our 6.5PN-accurate  new result 
\begin{widetext}
\begin{eqnarray}
\label{DeltaU_e2}
\delta U^{e^2}(u_p) &=& u_p+4 u_p^2+7 u_p^3+\left(-\frac{5}{3}-\frac{41}{32}\pi^2\right) u_p^4\nonumber\\
&+& \left(-\frac{11141}{45}+\frac{29665}{3072}\pi^2-\frac{592}{15}\gamma+\frac{3248}{15}\ln(2)-\frac{1458}{5}\ln(3)-\frac{296}{15}\ln(u_p)\right) u_p^5\nonumber\\
&+& \left(-\frac{2238629}{1575} +\frac{42282}{35}\ln(3)+\frac{8696}{105}\ln(u_p)+\frac{17392}{105}\gamma-\frac{167696}{105}\ln(2)-\frac{73145}{1536}\pi^2\right)u_p^6\nonumber\\
&-&  \frac{232618}{1575}\pi u_p^{13/2}\nonumber\\
&+& \left(\frac{2750367763}{198450} -\frac{13433142863}{3538944}\pi^2 +\frac{5102288}{2835}\gamma+\frac{41285072}{2835}\ln(2)-\frac{9765625}{4536}\ln(5)\right. \nonumber\\
&& \left. -\frac{673353}{280}\ln(3)+\frac{9735101}{262144}\pi^4+\frac{2551144}{2835}\ln(u_p)\right) u_p^7\nonumber\\
&+& \frac{2687231}{4410}\pi u_p^{15/2}+O(u_p^{8})\,.
\end{eqnarray}
\end{widetext}
We have also included in Eq. \eqref{deltaU_PN} the $O(e^4)$ contribution, $\delta U^{e^4}(u_p)$, obtained by using the recently derived 4PN EOB Hamiltonian \cite{Damour:2015isa}
together with the results of Ref. \cite{Tiec:2015cxa} (see below):
\begin{widetext}
\begin{eqnarray}
\label{DeltaU_e4}
\delta U^{e^4}(u_p) &=& -2 u_p^2+\frac14 u_p^3+\left(\frac{705}{8}-\frac{123}{256}\pi^2\right) u_p^4\nonumber\\
&+& \left(\frac{247931}{360}-\frac{89395}{6144}\pi^2+\frac{28431}{10}\ln(3)+\frac{292}{3}\gamma-\frac{64652}{15}\ln(2)+\frac{146}{3}\ln(u_p)\right)u_p^5+O(u_p^{11/2})\,,
\end{eqnarray}
\end{widetext}
as well as the 3PN-accurate $O(e^6)$ contribution \cite{Akcay:2015pza}
\begin{eqnarray}
\label{DeltaU_e6}
\delta U^{e^6}(u_p) &=& -\frac{5}{2} u_p^3+\left(-\frac{475}{12}+\frac{41}{128}\pi^2\right) u_p^4
\nonumber\\
&&
 +O(u_p^5)\,.
\end{eqnarray}

\section{Confirmation of recently derived 4PN results}

\vspace{7cm}

We are going to show that the 4PN-level restriction of our 6.5PN $O(e^2)$ result, Eq. \eqref{DeltaU_e2}, provides the first\footnote{Note, however, that the 4PN-level logarithmic terms in \cite{Damour:2015isa} agree with their previous determinations\cite{Damour:2009sm,Blanchet:2010zd,Barack:2010ny}.} independent analytical confirmation of the recently derived 4PN dynamics \cite{Jaranowski:2015lha,Bini:2013zaa,Damour:2014jta,Damour:2015isa}. In order to connect $\delta U(p,e)$ to the EOB formulation\cite{Damour:2014jta,Damour:2015isa} of the 4PN dynamics we make use of the recent results of Ref. \cite{Tiec:2015cxa}. The first step for making this connection is to transform the $e^2$-expansion of $\delta U$ into the corresponding $e^2$-expansion of $\delta z_1$. In view of the first  Eq. \eqref{eq_z1}, the coefficients of the $e^2$-expansion of $\delta z_1$,
\beq
\label{deltaz1_e_decomp}
\delta z_1=\delta z_1^{e^0}+e^2 \delta z_1^{e^2}+e^4 \delta z_1^{e^4}+O(e^6)\,,
\eeq
are, because of the $e^2$-dependence of  $U_0(p,e)$, linear combinations of several coefficients in the $e^2$-expansion of $\delta U$ (apart from the $O(e^0)$ Schwarzschild contribution which is simply $\delta z_1^{e^0}(u_p)=-(1-3u_p)\delta U^{e^0}(u_p)$).
Then, using  
\begin{widetext}
\begin{eqnarray}
\label{U_0_exp}
U_0(p,e) &\equiv& \frac{T_{r_0}}{{\mathcal T}_{r_0}}= \sqrt{\frac{p}{p-3}}\left(1-\frac32\frac{p^2-10p+22}{(p-2)(p-3)(p-6)}\, e^2  \right. \nonumber\\
&& -\frac38 \frac{(p^6-6 p^5-163 p^4+2188 p^3-10565 p^2+22860 p-18612)}{ (p-3)^2 (p-2)^3 (p-6)^3}\,  e^4 \nonumber\\
&& \left.  +\frac{1}{16}\frac{P_{11}}{(p-3)^3 (p-2)^5 (p-6)^5} e^6
+\frac{3}{1024}\frac{P_{16}}{(p-3)^4 (p-2)^7 (p-6)^7} e^8 
\right)+O(e^{10})\,,
\end{eqnarray}
where
\begin{eqnarray}
P_{11}&=& 6 p^{11}-275p^{10}+5606 p^9-67601 p^8+540759 p^7-3045312 p^6+12456657 p^5-37352007 p^4\nonumber\\
&& +80848488 p^3-120162744 p^2+109658448 p-46120752 \nonumber\\
P_{16}&=& 48 p^{16}-2992 p^{15}+87072 p^{14}-1573208 p^{13}+19787762 p^{12}-184077154 p^{11}+1313048541 p^{10}-7346722596 p^9\nonumber\\
&&+32702640748 p^8-116713090606 p^7+334571700617 p^6-766268642012 p^5+1380506243148 p^4\nonumber\\
&& -1895309547264 p^3+1868227475184 p^2-1176444492480 p+354281387328
\,,
\end{eqnarray}
we obtain
\begin{eqnarray}
\label{deltaz1_e2}
2 \delta z_1^{e^2}=&=&-2u_p+4 u_p^2+10u_p^3+\left(\frac{46}{3}+\frac{41}{16}\pi^2\right)u_p^4\nonumber\\
&+&
\left(\frac{20302}{45}-\frac{53281}{1536}\pi^2+\frac{1184}{15}\gamma-\frac{6496}{15}\ln(2)+\frac{2916}{5}\ln(3)+\frac{592}{15}\ln(u_p)\right)u_p^5\nonumber\\
&+& \left( -\frac{8704}{21}\gamma+\frac{504064}{105}\ln(2)-\frac{29160}{7}\ln(3)+\frac{246715}{1536}\pi^2+\frac{233158}{1575}-\frac{4352}{21}\ln(u_p) \right)u_p^6\nonumber\\
&+&\frac{465236}{1575}\pi u_p^{13/2}\nonumber\\
&+&\left(-\frac{8567728}{2835}\gamma-\frac{112700848}{2835}\ln(2)+\frac{1717281}{140}\ln(3)+\frac{9765625}{2268}\ln(5)+\frac{13871439695}{1769472}\pi^2\right. \nonumber\\
&& \left. -\frac{9735101}{131072}\pi^4-\frac{4750587838}{99225}-\frac{4283864}{2835}\ln(u_p)\right)u_p^7\nonumber\\
&-&\frac{4296083}{2205}\pi u_p^{15/2}\nonumber\\
&+& O(u_p^{8})\,.
\end{eqnarray}
and

\begin{eqnarray}
\label{deltaz1_e4}
\delta z_1^{e^4} &=& -u_p^2-\frac{19}4 u_p^3+\left(-\frac{339}{8}+\frac{123}{256}\pi^2\right)u_p^4\nonumber\\
&&+
\left(-\frac{31333}{180}+\frac{104155}{6144}\pi^2-\frac{28431}{10}\ln(3)-\frac{292}{3}\gamma+\frac{64652}{15}\ln(2)-\frac{146}{3}\ln(u_p)\right)u_p^5+O(u_p^6)\,.
\end{eqnarray}

\end{widetext}

Using Eq. (5.26) in Ref.  \cite{Tiec:2015cxa} (together with previous results connecting the main EOB radial potential to $\delta z_1^{e^0}$, see Refs. \cite{Barausse:2011dq, Akcay:2012ea}), we transformed the 6.5PN-accurate  knowledge of $\delta U^{e^2}$ \eqref{DeltaU_e2} into a corresponding 6.5PN-accurate knowledge of 
the second radial EOB potential $\bar D(u; \nu)$. More precisely, we found that the $O(\nu)$ contribution $\bar d(u)$ to the {\it function} $\bar D(u; \nu)=1+\nu \bar d(u)+O(\nu^2)$ is given by
\begin{widetext}
\begin{eqnarray}
\label{d_bar_eq}
\bar d(u)&=& 
6 u^2+52 u^3+\left(-\frac{533}{45}-\frac{23761}{1536}\pi^2+\frac{592}{15}\ln(u)-\frac{6496}{15}\ln(2)+\frac{1184}{15}\gamma+\frac{2916}{5}\ln(3)\right)u^4\nonumber\\
&&+\left(\frac{294464}{175}-\frac{63707}{512}\pi^2-\frac{1420}{7}\ln(u)+\frac{120648}{35}\ln(2)-\frac{2840}{7}\gamma-\frac{19683}{7}\ln(3)\right)u^5\nonumber\\
&&+\frac{264932}{1575}\pi u^{11/2}\nonumber\\
&& +\left(-\frac{64096}{45}\gamma-\frac{6381680}{189}\ln(2)+\frac{1765881}{140}\ln(3)+\frac{9765625}{2268}\ln(5)-\frac{31721400523}{2116800}+\frac{135909}{262144}\pi^4\right. \nonumber\\
&&\left. +\frac{229504763}{98304}\pi^2-\frac{32048}{45}\ln(u)\right) u^6\nonumber\\
&&-\frac{21288791}{17640}\pi u^{13/2}+O(u^7)\,.
\end{eqnarray}
\end{widetext}

Remarkably, the 4PN contribution to this so-calculated function, i.e., the (logarithmically-dependent) coefficient of $ u^4$
\begin{eqnarray}
&& -\frac{533}{45}-\frac{23761}{1536}\pi^2+\frac{592}{15}\ln(u)-\frac{6496}{15}\ln(2)\nonumber\\
&& +\frac{1184}{15}\gamma+\frac{2916}{5}\ln(3)
\end{eqnarray}
exactly coincides with the coefficient of $\nu u^4$ on the right-hand-side of Eq. (8.1b) in Ref. \cite{Damour:2015isa}.
As far as we know this is the first confirmation of the recently derived 4PN dynamics beyond the limit of circular orbits.

In addition, Ref. \cite{vandeMeent:2015lxa} (Table III) recently succeeded in extracting numerical estimates of the (4PN-level) coefficients of $e^2/p^5$ and $e^2 (\ln p)/p^5$ in $\delta U(p,e)$. The corresponding analytical result, i.e., the term of order  $O(u_p^5)$ in our result Eq. \eqref{DeltaU_e2}, reads
\begin{eqnarray}
&& \left(-\frac{11141}{45}+\frac{29665}{3072}\pi^2-\frac{592}{15}\gamma+\frac{3248}{15}\ln(2)\right. \nonumber\\
&& \left.-\frac{1458}{5}\ln(3)-\frac{296}{15}\ln(u_p)\right) u_p^5\,. 
\end{eqnarray}
Its numerical value is
\beq
\label{conf1}
(-345.3178497-19.73333333\ln(u_p)) u_p^5
\eeq
and this agrees, within the error bars, with the corresponding numerical estimates of Ref. \cite{vandeMeent:2015lxa} , namely 
\beq
\label{conf2}
(-345.37(5)-19.733(5)\ln u_p )u_p^5\,.
\eeq
Note that this additional agreement is a check both of the validity of the 4PN dynamics and of the relation   (5.26) in \cite{Tiec:2015cxa} [All the checks done in Ref. \cite{Tiec:2015cxa} were limited to the 3PN level]. 

Furthermore, we have displayed in Eq. \eqref{DeltaU_e4} above the analytical value of the coefficient of $e^4$ in $\delta U$, obtained by combining: 
(i)  relation (5.27) in \cite{Tiec:2015cxa}; 
(ii) the analytical 4PN result [derived in Eq. (8.1c) of  Ref. \cite{Damour:2015isa}] for the coefficient $q(u)$ of the contribution proportional to  $\nu  u^3 p_r^4 $ in the third EOB potential $Q(u, p_r; \nu)$;
(iii) our $\delta z_1^{e^2}$, Eq. \eqref{deltaz1_e2}, 
and (iv) the knowledge of $\delta z_1^{e^0}$.

The analytical  4PN-level contribution in $\delta U^{e^4}$, Eq.  \eqref{DeltaU_e4}, reads
\begin{eqnarray}
&&\left(\frac{247931}{360}-\frac{89395}{6144}\pi^2+\frac{28431}{10}\ln(3)+\frac{292}{3}\gamma\right. \nonumber\\
&& \left.-\frac{64652}{15}\ln(2)+\frac{146}{3}\ln(u_p)\right)u_p^5\,.
\end{eqnarray}
Its numerical value is 
\beq
\label{conf3}
\left(737.184955+48.66666667\, \ln(u_p)\right)u_p^5
\eeq
and this agrees, within the error bars, with the corresponding numerical estimates of Ref. \cite{vandeMeent:2015lxa} , namely
\beq
\label{conf4}
(737(4)+48.6(4)\ln u_p )u_p^5\,.
\eeq 
Again,  this further agreement is a check both of the validity of the 4PN dynamics and of the relations derived in     \cite{Tiec:2015cxa}. [Noticeably, the 4PN contribution to the EOB $q(u)$ potential does not involve $\ln(u)$. The corresponding logarithmic term $\frac{146}{3}\ln u$ in $\delta U^{e^4}(u)$ is generated during the transformation between $q(u)$ and $\delta U^{e^4}(u)$.]

Summarizing: among the four 4PN level coefficients related to non-circular dynamics ($d_4^c$, $d_4^{\ln{}}$, $q_3$, $q_6$) entering the EOB Hamiltonian derived in  Ref. \cite{Damour:2015isa} we have shown that the $O(\nu)$ contributions of {\it three} among them ($d_4^c$, $d_4^{\ln{}}$, $q_3$) agree either with the independent analytical calculations presented here or with recent SF-derived  numerical calculations.

\section{Confirmation of recently obtained 5PN and 5.5PN results}

The analytical 5PN-level contribution to $\delta U^{e^2}$ that we derived here reads
\begin{eqnarray}
&&\left(-\frac{2238629}{1575} +\frac{42282}{35}\ln(3)+\frac{8696}{105}\ln(u_p)+\frac{17392}{105}\gamma\right. \nonumber\\
&& \left.-\frac{167696}{105}\ln(2)-\frac{73145}{1536}\pi^2\right)u_p^6\,.
\end{eqnarray}
Its numerical value is
\beq
\label{conf5}
(-1575.580014+82.81904762\ln(u_p)) u_p^6\,.
\eeq
Ref. \cite{vandeMeent:2015lxa} (table III) recently succeeded in extracting numerical estimates of the (5PN-level) coefficients of $e^2/p^6$ and $e^2 (\ln p)/p^6$ in $\delta U(p,e)$. Their  estimates have large error bars and read
\beq
\label{conf6}
(-2000(400)+40(20)\ln(u_p)) u_p^6\,. 
\eeq
These estimates are compatible with our corresponding 5PN level results within \lq\lq one sigma" for the constant coefficient and within \lq\lq two sigma" for the (significantly smaller and less accurately determined) logarithmic coefficient.

Ref. \cite{Damour:2015isa}, generalizing the work of Ref. \cite{Bini:2013rfa} and using an effective-action approach, has shown 
that the second-order tail contribution to the two-body action (Eq. (9.19) in \cite{Damour:2015isa}) implied the existence of a  5.5PN-level term in the dynamics of eccentric binaries. In particular, they derived the following 5.5PN contribution  to the EOB $\bar D$ potential,
\beq
+\frac{264932}{1575}\pi \nu u^{11/2}\,.
\eeq 
This term agrees with our  independently derived 5.5PN contribution to $\bar d(u)$, Eq. \eqref{d_bar_eq}.
Let us note in passing that the high fractional errors in the estimates of the 5PN term in $\delta U^{e^2}$ of Ref. \cite{vandeMeent:2015lxa} might be linked to the non inclusion of a corresponding 5.5PN term $\propto u_p^{13/2}$ in $\delta U^{e^2}$.
{\it A contrario}, taking into account our new 6.5PN  analytical results might help in extracting more numerical information from existing SF numerical results on eccentric orbits.

\section{Comparison with  SF results on small-eccentricity orbits: $O(e^2)$-level}

Ref. \cite{Barack:2010ny}  succeeded in extracting (for the first time) gauge-invariant functional SF results for eccentric orbits by computing in the strong-field domain, $0< u \le \frac16$, the function  $\rho(u)$  parametrizing  the conservative $O(\nu)$ correction to the precession rate of small-eccentricity orbits.
Using the relation between $\rho(u)$ and the two $O(\nu)$ EOB potentials $a(u)$, $\bar d(u)$ derived in \cite{Damour:2009sm}, and the relation between $a(u)$ and $\delta z_1^{e^0}(u)$ \cite{Barausse:2011dq}, together with   accurate numerical calculations of $a(u)$ and $\delta z_1^{e^0}(u)$ in the strong-field domain, $0<u < \frac13$, Ref. \cite{Akcay:2012ea} computed the value of the function $\bar d(u)$ in the interval $0<u\le \frac16$ (see Table VI and Fig. 8 there). They also suggested that the function $\bar d(u)$ diverges at the light-ring $\propto (1-3u)^{-5/2}$.

\begin{figure}
\begin{center}
\[
\begin{array}{c}
\includegraphics[scale=0.30]{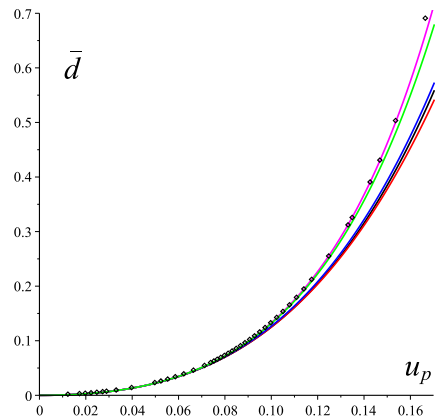}\cr
(a) \cr
 \includegraphics[scale=0.30]{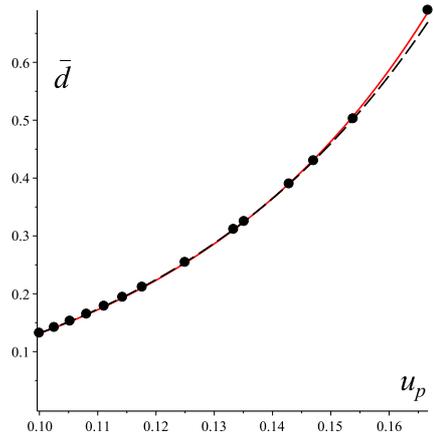} \cr
(b) \cr
\end{array}
\]
\caption{\label{fig:1} Panel (a): the successive PN-approximants to the function $\bar d(u_p)$ are compared with the SF numerical data  of Ref. \cite{Akcay:2012ea}.
Panel (b): the  numerical data  of  \cite{Akcay:2012ea} are confronted with two different analytical fits, 1) PN-like one given in Eqs. (9.40) and (9.41) of  Ref. \cite{Damour:2015isa} and 2) the Pad\'e-like fit of Eq. \eqref{bard_fit}, on the interval $0.1\le u_p \le 1/6$.}
\end{center}
\end{figure}

In the present work we succeeded in deriving the 6.5PN-accurate expansion of $\bar d(u)$, see Eq. (\ref{d_bar_eq}). 
In  panel (a) of Fig. 1 we study the convergence of the successive PN estimates towards the  SF numerical data  of Ref. \cite{Akcay:2012ea}.
Near the LSO they are ordered from bottom to top as: 5PN, 4PN, 5.5PN, 6.5PN, and 6PN.
Note that the best PN approximation is not provided by the formally most accurate 6.5PN one but by the previous one, i.e, by the 6PN approximant.[This is related  to the fact the 6.5 PN level contribution has a large and negative coefficient.]
Note in particular that the 6PN approximant predicts a value of $\bar d$ at the strongest field point $1/6$ (Last Stable Orbit, LSO) equal to $\bar d^{6PN}(1/6)\approx 0.664$, which is rather close to the numerical value $\bar d(1/6)=0.690(8)$\cite{Akcay:2012ea} and that, besides that point, its largest discrepancy with numerical data is $\approx +3 \times 10^{-3}$ at $u_p=1/7.4$.

In  panel (b) of Fig. 1 we compare (on the interval $1/10 \le u \le 1/6$) the  numerical data  of Ref. \cite{Akcay:2012ea} to two different analytical fits. One fit is the PN-like one given in Eqs. (9.40) and (9.41) of  Ref. \cite{Damour:2015isa}.  We derived the other one  by fitting to the data of \cite{Akcay:2012ea} a simple Pad\'e-like functional form  incorporating both some weak field information (first two PN terms) and the light-ring behavior of $\bar d(u)$ suggested in \cite{Akcay:2012ea}.
Our best-fit Pad\'e-like  representation of $\bar d(u)$ reads

\beq
\label{bard_fit}
\bar d^{\rm fit}(x)=6 x^2 \frac{(1+\frac{7}{6} x+5.2426\, x^2)}{ (1+30.2246\, x^2) (1-3 x)^{5/2} }\,.
\eeq
If we do not consider the LSO data point (which has a rather large  numerical uncertainty, $\sim 8\times 10^{-3}$),  
the   maximal difference of the Pad\'e-like fit, Eq. (\ref{bard_fit}), from the numerical data is about $5\times 10^{-3}$, while
the maximal difference from the numerical data of the PN-like fit \cite{Damour:2015isa} is about  $4\times 10^{-4}$. Though we think that the Pad\'e-like fit, Eq. \eqref{bard_fit} is probably a better global representation of $\bar d(u)$ in the full strong-field domain $0\le u\le \frac13$, we will use in the following the PN-like fit
because we shall only need an analytic representation of the function $\bar d(u)$ in the interval $0\le u\le \frac1{6.7}$.

We recall that the function $\bar d(u)$ is equivalent (via Eq. (5.26) of Ref. \cite{Tiec:2015cxa}) to the knowledge of $\delta U^{e^2}$ or $\delta z_1^{e^2}$, and therefore belongs to the $O(e^2)$-level  deviation from circularity.
Let us now compare the $ \sim 6$-digit accurate  calculations of $\delta U(p,e)$ of \cite{Barack:2011ed}  both to our high-order PN determination of $O(e^2)$ effects and our best-fit representation of the strong-field data on $\bar d(u)$ \cite{Akcay:2012ea}. In order to do this comparison we needed to extract from the sparse  numerical data on the function of two variables $\delta U(p,e)$ estimates of our theoretically convenient functions of only one variable $\delta U^{e^2}(u_p)$ and $\delta U^{e^4}(u_p)$, Eq. \eqref{deltaU_PN}.
Actually, we found it useful to work with the $e^2$ decomposition \eqref{deltaz1_e_decomp} of $\delta z_1(p,e)$ rather than
that of $\delta U(p,e)$.
Therefore, as a first step we converted the numerical data in Table IV of \cite{Barack:2011ed} into numerical data for $\delta z_1(p,e)$ (using Eq. \eqref{eq_z1} and the exact elliptic-integral value of $U_0(p,e)$). The result of this first step is displayed in Table I.

\begin{table}
\centering
\caption{Numerical values of $\delta z_1$ computed from \cite{Barack:2011ed}.}
\begin{ruledtabular}
\begin{tabular}{ccc}
$p$ & $e$ & $\delta z_1(p,e) $\cr
\hline
6.1& 0.021&   0.145665(1)\cr 
\hline
6.2& 0.05&  0.143680(1)\cr 
\hline
6.3& 0.1&   0.142549(1)\cr 
\hline
6.4& 0.1&   0.139272(1)\cr 
\hline
6.5& 0.1&  0.136575(1)\cr 
6.5& 0.2&    0.140003(1)\cr 
\hline
6.7& 0.1&  0.131928(1)\cr 
6.7& 0.2&   0.132593(1)\cr 
6.7& 0.3&    0.136057(1)\cr 
\hline
7& 0.1&    0.125950(1)\cr 
7& 0.2&       0.125240(1)\cr 
7& 0.3&        0.124298(1)\cr 
7& 0.4&       0.1240951(6)\cr 
7& 0.45&     0.1257752(5)\cr 
7& 0.49&     0.1331722(3)\cr 
7& 0.499&     0.1440447(2)\cr 
7& 0.4999&    0.15256636(2)\cr 
\hline
8& 0.3&      0.1052741(6)\cr 
8& 0.4&     0.1004073(7)\cr 
8& 0.5&      0.0936295(5)\cr
\hline 
9& 0.1&      0.0988833(7)\cr 
9& 0.2&      0.0967789(6)\cr 
9& 0.3&     0.0931712(5)\cr 
9& 0.4&      0.0878987(5)\cr 
9& 0.5&      0.0807111(4)\cr 
\hline
10& 0.1&     0.0896933(5)\cr 
10& 0.2&     0.0875910(5)\cr 
10& 0.3&   0.0840102(4)\cr 
10& 0.4&    0.0788294(4)\cr 
10& 0.5&   0.0718671(3)\cr 
\hline
15& 0.1&  0.06162446(8)\cr 
15& 0.2&    0.05994838(8) \cr 
15& 0.3&  0.05712804(8)\cr 
15& 0.4&   0.05312251(8)\cr 
15& 0.5&    0.04787370(8)\cr 
\hline
20& 0.1&   0.04701159(3)\cr 
20& 0.2&   0.04568137(3)\cr 
20& 0.3&   0.04345089(3)\cr 
20& 0.4&   0.04029993(3)\cr 
20& 0.5&  0.03620003(3)\cr
\end{tabular}
\end{ruledtabular}
\label{tab:1}
\end{table}

Among the data listed in Table I we could not make use of those providing  only one or two values of $e$ for a given value of $p$. This eliminates the data for $p=6.1,6.2,6.3,6.5$. In addition, we could not use the entry  $p=8$ because of the lack of data for $e=0.1$ and $e=0.2$ which  made it impossible for us to extract useful information.
For the other data, we extracted an estimate of $\delta z_1^{e^2}(u_p)$ by using only the three data points corresponding to $e=0.1,0.2,0.3$,  together with the value of $\delta z_1^{e^0}(p)=\delta z_1(p,e=0)$ encoded in the high-accuracy fit (model 14) provided in Ref. \cite{Akcay:2012ea}. We considered the subtracted and rescaled data
\beq
\label{widehatz1}
\widehat {\delta  z_1}\equiv \frac{\delta z_1(p,e)-\delta z_1(p,e=0)}{e^2}\,.
\eeq
Then we extracted two different estimates of $\delta z_1^{e^2}(u_p)$ from the latter data. The first estimate uses only the two points $e=0.1$ and $e=0.2$ and (uniquely) represents the two corresponding data as a linear function of $e^2$: $a+b e^2$. The second estimate uses   the three points $e=0.1$, $e=0.2$ and $e=0.3$ and  (uniquely) represents the three corresponding data as a quadratic  function of $e^2$: $a'+b' e^2+c' e^4$.
We then used: 1) the value of $a$ from the first operation as an estimate of $\delta z_1^{e^2}(u_p)$;  and 2) the difference $|a'-a|$ as an estimate of the error bar on $a$. 
The resulting numerically extracted estimates of $\delta z_1^{e^2}(u_p)$ (with their error bars) are displayed in the first column of Table II.

  
\begin{table}
\centering
\caption{Numerical/theoretical comparison for  $\delta z_1^{e^2}$.}
\begin{ruledtabular}
\begin{tabular}{cccc}
$p$ & $\delta z_1^{e^2\rm num}$& $\delta z_1^{e^2\rm th}{}_{{\rm mod}\#14, \bar d_{\rm fit}}$  & $\delta z_1^{e^2\rm PN}$\cr
\hline
20  & -0.044162(1) & -0.0441733       &-0.0441743   \cr    
15  & -0.055507(2) & -0.0555340  &-0.0555472    \cr     
10  & -0.069014(9) & -0.0691348   &-0.0696954   \cr    
9   & -0.06877(1)  & -0.0689361  &-0.0705279    \cr    
7   & -0.0252(1)   & -0.0255796   &-0.0538242    \cr    
6.7 & + 0.009(1)   & +0.00924715 &-0.0454867   \cr    
\end{tabular}
\end{ruledtabular}
\label{tab:2}
\end{table}

\begin{figure}
\begin{center} 
\includegraphics[scale=0.30]{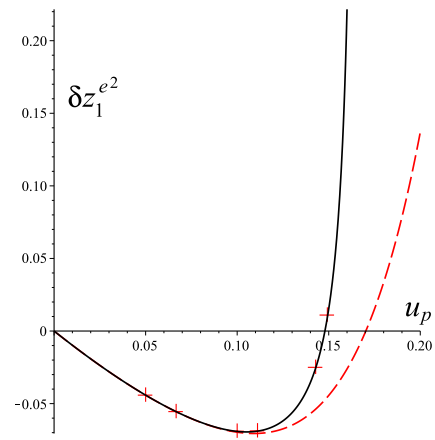}
\caption{\label{fig:2} The numerical SF data on $\delta z_1^{e^2}(u_p)$ (crosses, extracted by the procedure explained in the text) are compared with two theoretical models: 1) the expression of $\delta z_1^{e^2}(u_p)$ in terms of the SF-data-based analytical fits of the EOB potentials $a(u_p)$ and $\bar d(u_p)$ (solid curve) and 2) the 6.5PN expression of $\delta z_1^{e^2}(u_p)$ derived here (dashed curve). Error bars on the numerical data are too small to be visible on this scale.}
\end{center}
\end{figure}

These \lq\lq numerical" values are then compared to two different theoretical  estimates. The first theoretical estimate, displayed in the second column of Table II, was obtained by first using Eq. (5.26) of Ref. \cite{Tiec:2015cxa} to express $\delta z_1^{e^2}(u_p)$  in terms of the two EOB potentials $a(u)$ and $\bar d(u)$.
Then we replaced $a(u)$ by model 14 of Ref. \cite{Akcay:2012ea} and $\bar d(u)$ by the PN-like fit of Ref. \cite{Damour:2015isa}.
The second theoretical estimate, displayed in the third column of Table II, is the straightforward PN expansion of $\delta z_1^{e^2}(u_p)$ as given in Eq. \eqref{deltaz1_e2} above.
In addition, the comparison performed in Table II is visually represented in Fig. \ref{fig:2}.

The latter figure makes very clear two facts: (i) there is a good agreement between the numerically extracted $\delta z_1^{e^2}$ and the theoretical model incorporating both the theoretical link between EOB theory and  $\delta z_1^{e^2}$ and the current best SF-based representations of the two EOB potentials $a(u)$ and $\bar d(u)$;
(ii) though the 6.5PN-accurate expansion of $\delta z_1^{e^2}(u_p)$, Eq. \eqref{deltaz1_e2}, is in good agreement with the numerically extracted data for $u_p\lesssim 0.1$ ($p\gtrsim 10$), it fails to capture the numerical data as one approaches the LSO. Furthermore, it is interesting to note that our first theoretical estimate correctly predicts a change of sign of $\delta z_1^{e^2}(u_p)$ between $u_p=1/7$ and $u_p=1/6.7$. More precisely our first theoretical model predicts that $\delta z_1^{e^2}(u_p)$ should vanish at $u_p=1/p=1/6.760$. It would be interesting to check this prediction by doing SF simulations with $p=6.760$.

It is important to note that this change of sign close to the LSO is a simple consequence of the singular behavior of the function  $\delta z_1^{e^2}(u)$ near $u=1/6$.
Indeed, from Eq. (5.26) of Ref. \cite{Tiec:2015cxa} follows several facts. First,  $\delta z_1^{e^2}(u)$ can be expressed as the sum of three  contributions, namely
\beq
\label{z1_pieces}
\delta z_1^{e^2}(u)=\delta z_1^{e^2}{}_{\rm hom}(u)+\delta z_1^{e^2}{}_{a}(u)+\delta z_1^{e^2}{}_{\bar d}(u)\,.
\eeq 
Here the first \lq\lq homogeneous" contribution is defined as the expression that would remain  if $a(u)$ and $\bar d(u)$ were set to zero. The second term $\delta z_1^{e^2}{}_{a}(u)$ is a linear combination of $a(u)$ and its first two derivatives. Finally, the third term $\delta z_1^{e^2}{}_{\bar d}(u)$ is proportional to $\bar d(u)$.
It easily seen that $\delta z_1^{e^2}{}_{\bar d}(u)$ vanishes at the LSO proportionally to $(1-6u)\bar d(u)$.
The contribution $\delta z_1^{e^2}{}_{\rm hom}(u)$ is regular and nonvanishing at the LSO.  By contrast, the contribution $\delta z_1^{e^2}{}_{a}(u)$ diverges at the LSO
$\propto (1-6u)^{-1}$. [We are using here the fact that EOB theory predicts that the various EOB potentials are regular at the LSO. Their first singularity is located at the light-ring $u=1/3$ \cite{Akcay:2012ea}.] This means that we can theoretically predict the singular behavior at the LSO of the full $\delta z_1^{e^2}(u)$ from the sole knowledge of the main EOB potential $a(u)$. Using as above model 14 of Ref. \cite{Akcay:2012ea} we explicitly find the following singular behavior
\beq
\label{sing_behav1}
\delta z_1^{e^2\rm th}(u)= \frac{c_{-1}}{1-6u}+c_0+O(1-6u)\,,
\eeq
with the following numerical values 
\beq
\label{sing_behav2}
c_{-1}\simeq + 0.0136455 \,\qquad c_0 \simeq-0.116733\,.
\eeq
The fact that $c_{-1}$ is positive then predicts that $\delta z_1^{e^2}(u)=-u+O(u^2)$ which is negative in the weak-field domain ($u\ll 1$)
must change sign before reaching the LSO, thereby explaining the change of sign found above.
Let us mention the simple link existing between $\delta z_1^{e^2}(u)$ and the function $\rho(u)$ (introduced in \cite{Damour:2009sm}) measuring the precesssion of small eccentricity orbits. Eliminating $\bar d(u)$ between Eq. (36) and the similar expression, derived in \cite{Damour:2009sm}, linking $\rho(u)$ to $\bar d(u)$, $a(u)$, $a'(u)$ and $a''(u)$ we find
\begin{eqnarray}
\label{rho_TD}
\rho(u)&=&4u+ 2\frac{1-2u}{u}\sqrt{1-3u}\delta z_1^{e^2}(u)\nonumber\\
&& +2 (1-10u+22u^2)\sqrt{1-3u} \times \nonumber\\
&& \left[\frac{1}{1-6u}\frac{d}{du}\left(\frac{a(u)}{\sqrt{1-3u}}  \right)+\frac{1-2u}{(1-3u)^2}  \right]\,.
\end{eqnarray}
As $\rho(u)$ is a regular function near the LSO this relation shows that the origin of the LSO-singular behavior of $\delta z_1^{e^2}(u)$ is the term 
\begin{eqnarray}
\label{eq_z1_LSO}
\delta z_1^{e^2}(u)&=&-\frac{u(1-10u+22u^2)}{(1-2u)(1-6u)}\frac{d}{du}\left(\frac{a(u)}{\sqrt{1-3u}}  \right)\nonumber\\
&& +\hbox{\rm LSO-regular}\,.
\end{eqnarray}

\section{Going beyond the  $O(e^2)$-level}

\subsection{Comparison with  $O(e^4)$ information extracted from SF results}

We have indicated above how we extracted the $O(e^2)$ contribution  $\delta z_1^{e^2}(u)$ to $\delta z_1(p,e)$ from a part of the data listed in Table IV of \cite{Barack:2011ed}.
The procedure we used, based on representing the subtracted and rescaled data $\widehat {\delta z_1}$, Eq. \eqref{widehatz1}, either as $a+b e^2$ or $a'+b'e^2 +c' e^4$ gives also an estimate of 
 the $O(e^4)$ contribution $\delta z_1^{e^4}(u)$ to $\delta z_1(p,e)$, namely the value of $b$. In addition, the difference $|b'-b|$ gives an estimate of the error bar on $\delta z_1^{e^4}(u)$.  
The resulting numerically extracted estimates of $\delta z_1^{e^4}(u_p)$ (with their error bars) are displayed in the first column of Table III.

\begin{table}
\centering
\caption{Numerical/theoretical comparison for  $\delta z_1^{e^4}$.}
\begin{ruledtabular}
\begin{tabular}{cccc}
$p$ & $\delta z_1^{e^4\rm num}$&  $\delta z_1^{e^4\rm PN}$\cr
\hline
20 &-0.0036(1)  &     -0.00334554  \cr      
15 & -0.0072(3) &     -0.00668353  \cr       
10 & -0.021(1)  &     -0.0193812  \cr       
9 & -0.027(1) &       -0.0261540 \cr       
7 & +0.03(2) &        -0.0561272  \cr      
6.7 & +0.3(2)  &      -0.0646171 \cr       
\end{tabular}
\end{ruledtabular}
\label{tab:3}
\end{table}

In the second column of the latter table we compare the so extracted numerical estimates to the values of $\delta z_1^{e^4}(u_p)$ predicted by the straightforward PN expansion, Eq. \eqref{deltaz1_e4}, deduced from the 4PN knowledge of $q(u)$, \cite{Damour:2015isa}, together with the results of Ref. \cite{Tiec:2015cxa}. [Because of the four derivatives of $a(u)$ entering Eq. (5.27) there we found that the use of model 14 leads to inaccuracies too large for getting reliable results.] It is satisfactory
to notice that the theoretical estimates are compatible within about twice the indicated error bars for all points except for the last two (near LSO) ones.
This indicates that with the present data it seems rather difficult to extract accurate strong-field information going beyond the current theoretical knowledge of $\delta z_1^{e^4}(u_p)$.

\subsection{Comparison with  SF results on eccentric orbits}

\begin{figure}
\begin{center} 
\includegraphics[scale=0.30]{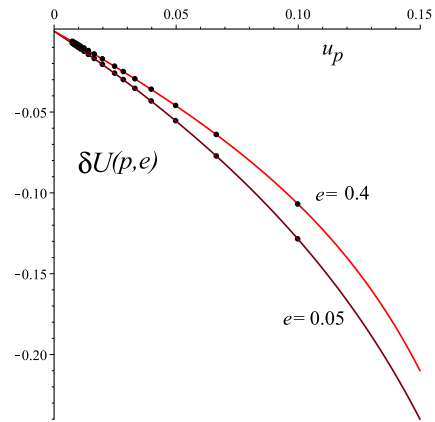}
\caption{\label{fig:3}Numerical SF data points (from Ref.\cite{Akcay:2015pza}) for  $\delta U(p,e)$ are compared with the sum of $\delta U(p,0)$ (given by model 14 in \cite{Akcay:2012ea}) and of the  PN-expanded analytical prediction of Eqs.
\eqref{DeltaU_e2}, \eqref{DeltaU_e4}, \eqref{DeltaU_e6}. 
We consider the two extreme eccentricities listed in Table II of  \cite{Akcay:2015pza}, namely  $e=0.05$  and $e=0.4$.}
\end{center}
\end{figure}

In an attempt to bypass the difficulty of decomposing the numerical function $\delta z_1(p,e)$  (or for that matter $\delta U(p,e)$) into various powers of $e^2$
 we also performed direct comparisons between numerical data on $\delta U(p,e)$ and the combined theoretical result obtained by summing: (i) model 14 for $\delta U(p,e=0)$; (ii) our 6.5PN-accurate result, Eq. \eqref{DeltaU_e2} for the $e^2$ contribution, (iii) the 4PN-accurate result, Eq. \eqref{DeltaU_e4}, deduced above 
and (iv) the 3PN terms  for the  $e^6$ contribution given in Eq. (4.53d) of Ref. \cite{Akcay:2015pza}. 
Such a comparison is done in Fig. \ref{fig:3} using as numerical data points a sample of the SF data recently computed in Ref. \cite{vandeMeent:2015lxa}.
The agreement exhibited in Fig. \ref{fig:3} is rather satisfactory and confirms the difficulty in extracting from numerical data information beyond the current theoretical knowledge.

\section{Conclusions}

Let us summarize our main results.

The gauge-invariant self-force $O(m_1/m_2)$ correction $\delta U$ to the averaged inverse redshift function $U(m_2\Omega_r, m_2\Omega_\phi, m_1/m_2)=\oint dt/\oint d\tau$ along an eccentric orbit around a Schwarzschild black hole can be viewed as a function of the (dimensionless) semi-latus rectum $p$ and the eccentricity $e$ of the (unperturbed) orbit. The function $\delta U(p,e)$ can be expanded in powers of $e^2$: $\delta U(p,e)=\delta U^{e^0}(u_p)+e^2 \delta U^{e^2}(u_p)+e^4 \delta U^{e^4}(u_p)+e^6 \delta U^{e^6}(u_p)+\ldots$, where $u_p=1/p$.
We computed, by a direct analytic self-force computation along slightly eccentric orbits, the PN-expansion of the term $\delta U^{e^2}(u_p)$ up to  order $O(u_p^{15/2})$ included (corresponding to the 6.5PN level), see Eq. \eqref{DeltaU_e2}. We completed this result by giving the 4PN-accurate expansion of $\delta U^{e^4}(u_p)$, Eq. \eqref{DeltaU_e4}, deduced by inserting the 4PN Hamiltonian of \cite{Damour:2015isa} into the relations recently derived in \cite{Tiec:2015cxa}. [The present knowledge of the next term $\delta U^{e^6}(u_p)$ is limited at the 3PN-level, Eq. \eqref{DeltaU_e4}, see Ref. \cite{Akcay:2015pza}.]
We gave the corresponding results for the self-force correction $\delta z_1=-U^{-2}\delta U$ to the averaged redshift $z_1=1/U=\oint d\tau/\oint dt$, see Eqs. \eqref{deltaz1_e2} and \eqref{deltaz1_e4}.

Using the relations derived in \cite{Tiec:2015cxa}, we converted our 6.5PN expansion of $\delta U^{e^2}$ into the corresponding 6.5PN-accurate expansion of the $O(\nu)$ contribution to $\bar d(u)$ to the second radial EOB potential $\bar D(u; \nu)=1+\nu \bar d(u)+O(\nu^2)$ (which enters the dynamics of eccentric  orbits at the $p_r^2$ level), see Eq. \eqref{d_bar_eq}.

The 4PN-level comparison between the latter result and the recently derived 4PN-accurate EOB Hamiltonian \cite{Damour:2015isa}, has given us the first independent analytic confirmation of the 4PN dynamics beyond the limit of circular orbits. We also showed that recent numerical computations of self-force effects along eccentric orbits \cite{vandeMeent:2015lxa} gave two more (numerical) confirmations of the 4PN dynamics, at the $O(e^2)$, and at the $O(e^4)$ levels, see Eqs. \eqref{conf1} and \eqref{conf2} and Eqs. \eqref{conf3} and \eqref{conf4}. The same numerical computations gave a further rough confirmation of the 5PN contribution to our 6.5PN $O(e^2)$ result, see Eqs. \eqref{conf5} and \eqref{conf6}.
Finally, we pointed out that our result has also confirmed the recent calculation of the 5.5PN contribution to the $O(e^2)$ dynamics achieved in \cite{Damour:2015isa}.

In addition to confirming and extending various post-Newtonian and effective-one-body results describing the dynamics of eccentric orbits, we have directly compared our
high-order analytic results to various numerical calculations of self-force effects in eccentric orbits \cite{Barack:2010ny,Barack:2011ed,Akcay:2012ea,vandeMeent:2015lxa,Akcay:2015pza}. The results of our comparisons are displayed in Figs.  \ref{fig:1}, \ref{fig:2} and \ref{fig:3}  and in Tables II and III. 
This comparison shows that our high PN order results accurately agree with numerical results up to gravitational potentials $u\le 0.1$ (corresponding to semi-latus recta $p\le 10$). On the other hand, in the strong-field domain $0.1\simeq u \simeq \frac16$, a good agreement with the recent  eccentric redshift self-force data is reached (as illustrated in Fig. \ref{fig:2}) only if one replaces the current 6.5PN expanded analytic knowledge of $O(e^2)$ effects by combining EOB theory (which describes $O(e^2)$ effects by the secondary potential $\bar d(u)$) with analytic fits of the EOB potential $\bar d(u)$ obtained from {\it previous} numerical self-force data on the precession of small-eccentricity orbits \cite{Barack:2010ny,Akcay:2012ea}. 

We hope that our new, analytic 6.5PN $O(e^2)$ results will help to extract more information from numerical self-force calculations both at the order $O(e^2)$ and at higher orders in $e^2$. From the point of view of EOB theory (and of its application to comparable-mass binaries) it would be most useful to extract information about the third $O(\nu)$ EOB radial potential (beyond $a(u)$ and $\bar d(u)$), namely the function $q(u)$ entering $O(e^4)$ effects. The preliminary analysis we presented in Section VI A indicates what is needed for this. One would need a denser set of dedicated self-force computations containing, for various values of $u_p=1/p$ uniformly (except for an increased density near $1/6$) covering the interval $0\le u_p\le 1/6$, a set of small-enough eccentricity values able to accurately extract the coefficient of the $O(e^4)$ contribution to $\delta z_1(p,e)$.

In this respect, let us end by commenting on the analytic structure of the function $\delta z_1(p,e)$. Note, first, that when expanding $\delta z_1 (p,e)$ in powers of $e^2$, the coefficients of the successive powers of $e^2$ have an increasingly {\it singular} behavior near the last stable orbit (LSO) at $p=6$.
To start with, $\delta z_1^{e^0}(p)=\delta z_1 (p, e=0)$ is regular at the LSO (as follows, say, from its EOB link with the first $O(\nu)$ EOB potential $a(u)$ whose first singularity is at the light-ring \cite{Akcay:2012ea}. Then, the $O(e^2)$ piece $\delta z_1^{e^2}(u_p)$ has a $\sim 1/(1-6u_p)$ singularity at the LSO. We numerically computed in Eqs. \eqref{sing_behav1} and \eqref{sing_behav2} the values of the first two coefficients of the Laurent expansion of $\delta z_1^{e^2}(u)$ deduced from the knowledge of the EOB $O(\nu)$ potential $a(u)$. [We note in passing that it would be interesting to numerically check the predictions \eqref{sing_behav1} and \eqref{sing_behav2} as well as the value $p_0=6.760$, we estimated, where $\delta z_1^{e^2}(1/p)$ vanishes before growing towards $+\infty$ as $p\to 6$.]  
The corresponding Laurent expansion of the $O(e^2)$ piece $\delta U^{e^2}(u)$ in $\delta U(p,e)$ reads
\beq
\delta U^{e^2}(u) \simeq -\frac{0.05196}{1-6u}+0.5717\,.
\eeq
Note that the presence of a $\sim 1/(1-6u)$ singularity in $\delta z_1^{e^2}(u)$ is particularly clear from the link Eq. \eqref{rho_TD} between the $O(e^2)$ precession function $\rho(u)$ (introduced in \cite{Damour:2009sm} and defined there so as to be regular across the LSO), $\delta z_1^{e^2}(u)$ and $a(u)$: see Eq. \eqref{eq_z1_LSO}.

The $O(e^4)$ piece $\delta z_1^{e^4}(u)$ has, as a consequence of Eqs. (5.26) and (5.27) in \cite{Tiec:2015cxa}, a link to the first three $O(\nu)$ EOB potentials of the type
\beq
\label{ze4q_var_pieces}
\delta z_1^{e^4}(u)=\delta z_1^{e^4}{}_{\rm hom}(u)+\delta z_1^{e^4}{}_{a}(u)+\delta z_1^{e^4}{}_{\bar d}(u)+\delta z_1^{e^4}{}_{q}(u)\,.
\eeq 
Note that, as in Eq. \eqref{z1_pieces}, the first \lq\lq homogeneous" contribution is analytically known (and LSO-regular); the second term is a linear combination of $a(u)$ and its first four derivatives; the third term is a linear combination of $\bar d(u)$ and its first two derivatives; while the fourth and last term is proportional to $q(u)$ and explicitly given by
\beq
\label{ze4q}
\delta z_1^{e^4}(u)=9 \frac{u^2(1-6u)^2}{(1-2u)^4(1-3u)^{3/2}} q(u)\,.
\eeq
We only know the 4PN level expansion of the EOB potential $q(u)$ \cite{Damour:2015isa}, $q(u)=c_2u^2+c_3u^3$ (see Eq. (8.1c) in \cite{Damour:2015isa}) and, in particular we do not know the value of $q(u)$ at the LSO (besides the fact that the EOB theory predicts that $q(u)$ is regular near the LSO). We see, however,  from Eq. \eqref{ze4q}, that, near the LSO, the effect of $q(u)$ is $O((1-6u)^2)$. On the other hand, the terms $\delta z_1^{e^4}{}_a$ and $\delta z_1^{e^4}{}_{\bar d}$ involve LSO-singular terms of the symbolic type (indicating only the power of the LSO singulariy)
\begin{eqnarray}
\delta z_1^{e^4}{}_a &\sim & \frac{a(u)}{(1-6u)^{3}}+\frac{a'(u)}{(1-6u)^{3}}+\frac{a''(u)}{(1-6u)^{2}}\nonumber\\
&+& \frac{a'''(u)}{(1-6u)}\nonumber\\
\delta z_1^{e^4}{}_{\bar d}&\sim & \frac{\bar d(u)}{(1-6u)}\,.
\end{eqnarray}
Using the model 14 analytic fit of $a(u)$  \cite{Akcay:2012ea} together with our Pad\'e like fit, Eq. \eqref{bard_fit}, for $\bar d(u)$ (which, according to Fig. 1 b seems to better capture the LSO behavior of $\bar d(u)$), we deduce, from Eqs. \eqref{ze4q_var_pieces} and \eqref{ze4q},  that the theoretically predicted LSO behavior of $\delta z_1^{e^4}(u)$ is of the type
\begin{eqnarray}
\delta z_1^{e^4}(u)&=& \frac{c'_{-3}}{(1-6u)^3}+\frac{c'_{-2}}{(1-6u)^2}+\frac{c'_{-1}}{(1-6u)}+c'_0\nonumber\\
&+& c'_1(1-6u)+O((1-6u)^2)\,,
\end{eqnarray}
with the following approximate numerical values for the various coefficients of the Laurent expansion
\begin{eqnarray}
c'_{-3}&\simeq & +0.0004264\nonumber\\
c'_{-2}&\simeq&-0.001279 \nonumber\\
c'_{-1}&\simeq&+0.0006447 \nonumber\\
c'_0&\simeq&-0.09396 \nonumber\\
c'_1&\simeq& +0.3435\,.
\end{eqnarray}
It would be interesting to confirm this prediction by means of dedicated, near LSO, numerical self-force computations.

Summing the various pieces of $\delta z_1(p,e)=\delta z_1^{e^0}(p)+e^2 \delta z_1^{e^2}(p)+\ldots$ the structure of the $e^2$-dependence of the LSO-singular behavior of
$\delta z_1(p,e)$ is essentially of the type
\beq
\delta z_1(p,e)\sim 1 +\frac{e^2}{p-6}+\frac{e^4}{(p-6)^3}+O(e^6)\,.
\eeq
The reason why the $O(e^4)$ term is more LSO-singular than the square of the $O(e^2)$ term, is that one should understand this structure as being of the type
\begin{eqnarray}
\label{new_d_z1}
\delta z_1(p,e)&=&\delta z_1(p,0)+\frac{4e^2}{p-6}f_p(\epsilon)\nonumber\\
&=& \delta z_1(p,0)+(p-6) \epsilon f_p(\epsilon)
\end{eqnarray}
where
\beq
\label{epsilon}
\epsilon\equiv \left(\frac{2e}{p-6} \right)^2\equiv \left( \frac{e}{e_c(p)} \right)^2
\eeq
and where
\beq
\label{fepsilon}
f_p(\epsilon)=f_0(p)+f_1(p) \epsilon +f_2(p) \epsilon^2 +\ldots + f_{2n}\epsilon^{2n}
\eeq
is a function of $\epsilon$ which is {\it analytic} near $\epsilon=0$.
[Here, and henceforth, we focus on the $\epsilon$-dependence of $f_p(\epsilon)$, which, after the factorizations done in Eq. \eqref{new_d_z1} should be a regular function of $p$ near the LSO]. Indeed, $z_1(m_2\Omega_r, m_2\Omega_\phi)$  (and therefore $\delta z_1(p,e)$) is an even \footnote{This is seen if we think of $z_1$ as a function of $E^2$ and $L^2$ via Eqs. \eqref{E_and_L_in_p_e}} function of $e$, whose singularities in the complex $e$-plane lie on the two boundaries of the inequalities \eqref{separatrix}, namely at $e=e_c(p)\equiv \frac{p-6}{2}$ and $e=1$. For discussing the singularity structure of the small-$e^2$ expansion of $\delta z_1(p,e)$, it is the first, \lq\lq separatrix" boundary, $e=\frac{p-6}{2}$ which matters. This boundary corresponds to a locus where the definition of the function $z_1(p,e;\lambda)$, with $\lambda=m_1/m_2$ [here considered as a function of some exactly defined versions of $p$ and $e$, say through (gauge-invariant) EOB theory] breaks down because of the disappearance of  stable bound orbits oscillating between $r_{\rm peri}=Mp/(1+e)$ and $r_{\rm apo}=Mp/(1-e)$.
In general mathematical terms, one can view the function $z_1(p,e;\lambda)$ as a \lq\lq period" \cite{kont} over a cycle. This period becomes singular (as well as its derivative with respect to the small deformation parameter $\lambda=m_1/m_2$) when the cycle ceases to exist, or abruptly changes character.
For a given value of $p$, the change of nature of the cycle (between  $r_{\rm peri}$ and  $r_{\rm apo}$) happens at min$(e_c(p),1)$. When $p<8$, one first encounters the singularity at $e=e_c(p)=(p-6)/2$. Viewing the function $f(\epsilon)$ in Eq. \eqref{new_d_z1} as an analytic function in the complex $\epsilon$-plane, the location of the first singularity determines the radius of convergence of its Taylor expansion around $\epsilon=0$. In view of the definition, Eq. \eqref{epsilon},  of $\epsilon$ this singularity (if $p<8$) is located at $\epsilon=1$. We therefore expect the expansion \eqref{fepsilon}
to have a radius of convergence equal to 1, i.e., that the rescaled expansion coefficients $f_n/f_0$ are (roughly) of order unity. These considerations give a guideline for choosing, for each value of $p$, the value of $e$ one should explore. Essentially, one wants (at least to explore the near-LSO region $6<p<8$) to have a sample of values of $\epsilon$ (with $0<\epsilon<1$) which is sufficiently dense and uniform (and sufficiently close to $0$) to be able to numerically extract the values of the expansion coefficients $f_0(p)$, $f_1(p)$, ....
The coefficient $f_0(p)$ parametrizes $\delta z_1^{e^2}(p)$ (and therefore $\bar d(u_p)$), while the coefficient $f_1(p)$ parametrizes $\delta z_1^{e^4}(p)$ (and therefore $q(u_p)$), etc. Such a procedure might help to extract the strong-field behavior of the EOB potential $q(u)$. [Actually, as discussed in \cite{Damour:2015isa}, $q(u)=q_4(u)$ is only the first element in a sequence $q_4(u)$, $q_6(u)$, $q_8(u)$,... parametrizing the coefficients of $p_r^4$, $p_r^6$, $p_r^8$, .... This sequence is in correspondence with $f_0(u_p)$, $f_1(u_p)$, $f_2(u_p)$,...]

To conclude, let us emphasize that studies of the $e^2$-expansion of $\delta z_1(p,e)$ have the defect of being able to explore the two-body dynamics behind it (say in its EOB formulation to be concrete) {\it only} in the medium-strong-field domain $0<u<\frac16$. It cannot access the really strong-field domain $\frac16 \le u <\frac13$ where the various $O(\nu)$ EOB potentials are a priori defined and regular.
In view of this limitation, we recommend that the self-force community make an effort to implement the suggestions made in Ref. \cite{Damour:2009sm}. Indeed, \cite{Damour:2009sm} (notably see Sec. VI there) suggested several different ways of extracting {\it gauge-invariant} information from self-force theory that might be useful for informing the dynamics of comparable-mass binaries (notably in its EOB formulation). In particular,  \cite{Damour:2009sm} suggested to compute the gauge-invariant functional link $\theta({\mathcal E},J)$ between the (total, conserved) energy ${\mathcal E}$ and angular momentum $J$ and the scattering angle $\theta$ of hyperbolic-like orbits. To avoid having to correct for the effect of the radiation-damping part of the self-force (though an appropriate method for doing so was provided in \cite{Bini:2012ji}) it would be best to compute the function $\theta_{\rm cons}({\mathcal E},J)$ associated with the {\it conservative} part of the self-force. As mentioned in \cite{Damour:2009sm}, the function of two variables  $\theta_{\rm cons}({\mathcal E},J)$ contains \lq\lq ample information for determining the functions entering the EOB formalism."
We note in particular here that this function has the potential of probing the functions $\bar d(u)$, $q_{2n}(u)$ in the full strong-field domain $0<u<\frac13$.
This information would usefully complement the recent work \cite{Damour:2014afa} which succeeded in probing the dynamics of comparable-mass binaries by extracting $\theta({\mathcal E},J)$ from full numerical relativity simulations of hyperbolic-like close binary black hole encounters.

\subsection*{Acknowledgments}
We thank Seth Hopper for attracting our attention to Ref. \cite{vandeMeent:2015lxa}.
D.B. thanks the Italian INFN (Naples) for partial support and IHES for hospitality during the development of this project.
All  the authors are grateful to ICRANet for partial support.

\end{document}